\documentclass[english,twocolumn,showpacs,preprintnumbers,amsmath,amssymb,superscriptaddress,pra]{revtex4-1}
\usepackage{mathptmx}
\usepackage[T1]{fontenc}
\usepackage[latin9]{inputenc}
\usepackage{amsmath}
\usepackage{amssymb}
\usepackage{graphicx}

\makeatletter
 
 \@ifundefined{textcolor}{}
 {%
   \definecolor{BLACK}{gray}{0}
   \definecolor{WHITE}{gray}{1}
   \definecolor{RED}{rgb}{1,0,0}
   \definecolor{GREEN}{rgb}{0,1,0}
   \definecolor{BLUE}{rgb}{0,0,1}
   \definecolor{CYAN}{cmyk}{1,0,0,0}
   \definecolor{MAGENTA}{cmyk}{0,1,0,0}
   \definecolor{YELLOW}{cmyk}{0,0,1,0}
 }

\makeatother

\usepackage{babel}
\begin{document}

\title{Efficient frequency conversion through absorptive bands of the nonlinear
crystal}

\author{Gil Porat}

\email{gilpor@gmail.com}

\selectlanguage{english}%

\author{Ady Arie}

\affiliation{Department of Physical Electronics, Fleischman Faculty of Engineering,
Tel Aviv University, Tel Aviv 69978, Israel}
\begin{abstract}
Two simultaneous three wave mixing processes are analyzed, where an
input frequency is converted to an output frequency via an intermediate
stage. By employing simultaneous phase-matching and an adiabatic modulation
of the nonlinear coupling strengths, the intermediate frequency is
kept dark throughout the interaction, while obtaining high conversion
efficiency. This feat is accomplished in a manner analogous to population
transfer in atomic stimulated Raman adiabatic passage (STIRAP). Applications
include conversion between remote frequencies, e.g. mid-IR to visible,
and study of electronic crystal properties in the UV absorption band.
\end{abstract}


\maketitle

\section{Introduction}

Quadratic nonlinear optical media are commonly used to perform optical
frequency conversion via three wave mixing (TWM) processes \cite{Boyd_NLO_book}.
In this way, laser frequencies which are not available by direct laser
action are generated. Combinations of such processes were utilized
extensively in order to reach frequencies either far removed from
a laser source or close to it \cite{Saltiel_PO_47,Tehranchi_OE_19}.
This task is not trivial, since in order for a TWM process to be efficient
a phase-matching condition has to be fulfilled. For a single TWM process,
a well known solution is quasi phase matching (QPM) \cite{Hum_CRP_8},
e.g. by periodically modulating the sign of the nonlinear coefficient.
Schemes also exist for simultaneous phase-matching of multiple processes
\cite{Lifshitz_PRL_95,Chou_OL_24}, at the expense of conversion efficiency.
Furthermore, the medium is required to be transparent at all participating
frequencies, including any frequency generated as an intermediate
stage before further conversion.

Recently, analogies were shown between the dynamics of simultaneous
TWM processes and those induced in three level atoms by electromagnetic
fields. In one publication, an analogy is utilized to perform second
harmonic generation and sum frequency generation (SFG) simultaneously,
such that the third harmonic is generated efficiently \cite{Longhi_OL_32}.
In this case, the dynamics is nonlinear and significant power was
generated at the intermediate frequency (second harmonic). In another
work the analogy was applied to the case of purely linear dynamics,
where two simultaneous sum or difference generation processes (DFG),
or a combination of them, were performed under the undepleted pump
approximation \cite{Porat_OE_20}. With this method, called adiabatic
elimination, only negligible power is ever present at the intermediate
frequency throughout the entire conversion process, so transparency
at the intermediate frequency is not required. However, this method
explicitly requires that each one of the two processes has a large
phase-mismatch. Due to inherently large phase-mismatches, high conversion
efficiency is difficult to achieve, and requires high pump intensities
or very long interaction lengths.

Here, we propose a method to perform two simultaneous TWM processes
efficiently and without significant generation of an intermediate
frequency. In contrast to adiabatic elimination \cite{Porat_OE_20},
the two processes are phase-matched throughout the entire interaction
length. High efficiency can therefore be maintained even in the presence
of significant absorption of the intermediate frequency wave. As a
result, one of the nonlinearity enhancement mechanisms which is associated
with absorption resonances can be exploited. In this case, the analogous
atomic scenario is the so-called stimulated Raman adiabatic passage
(STIRAP) \cite{Tannor_QM_book,Vitanov_ARPC_52}. We emphasize that
here the dynamics are linear, in complete correspondence to STIRAP.
It is this linearity which enables the elimination of the intermediate
frequency, in contrast to the previous work by Longhi \cite{Longhi_OL_32},
which mimicked a nonlinear version of STIRAP, and consequently significant
intermediate power was present ($\sim20\%$ of the input power).

The insensitivity to intermediate frequency absorption makes possible
processes in which the intermediate frequency is in the ultraviolet
(UV) absorption band of a nonlinear crystal. This method can thus
be used to probe the electronic properties of the crystal in this
spectral region, a task which was previously made difficult by absorption.

This paper is organized as follows. In section \ref{sec:Theoretical-Model-and},
the theoretical model is presented and adiabatic interaction is analyzed.
In section \ref{sec:Numerical-Simulation-Ideal}, two study cases,
which demonstrate the main features of our method, are simulated numerically
assuming ideal conditions. Section \ref{sec:Numerical-Simulation-PRQPM}
repeats the calculation of section \ref{sec:Numerical-Simulation-Ideal}
under conditions that are attainable with current QPM technology,
and discusses its limitations. In section \ref{sec:Future-Prospects-and}
we discuss further improvements of the QPM-based method and another
way of achieving nonlinear optics STIRAP, as well as applications
of this technique.

\section{Theoretical Model and Analysis\label{sec:Theoretical-Model-and}}

\subsection{Theoretical model and atomic analogy}

In this model, two simultaneous TWM processes are considered, such
that the frequency generated by one process is further combined with
another frequency in the other process. For example, a SFG process
$\omega_{2}=\omega_{1}+\omega_{p_{1}}$ can be accompanied by a DFG
process $\omega_{3}=\omega_{2}-\omega_{p_{2}}=\omega_{1}+\omega_{p_{1}}-\omega_{p_{2}}$,
where $\omega_{1}$ is the input frequency, $\omega_{2}$ is the intermediate
frequency, $\omega_{3}$ is the output frequency and $\omega_{p_{1}}$and
$\omega_{p_{2}}$ are the first and second pump frequencies, respectively.
We make the undepleted pump approximation: the two pump waves are
taken to be much more intense than the other waves, thus they are
negligibly affected by the interaction. Furthermore, all beams are
assumed to be plane waves. If one uses Gaussian beams, this requirement
can be translated to a Rayleigh range $z_{R}$ that satisfies $z_{R}\gtrsim L/2$,
where $L$ is the nonlinear medium's length.

\begin{figure}
\begin{centering}
\includegraphics[width=1\columnwidth]{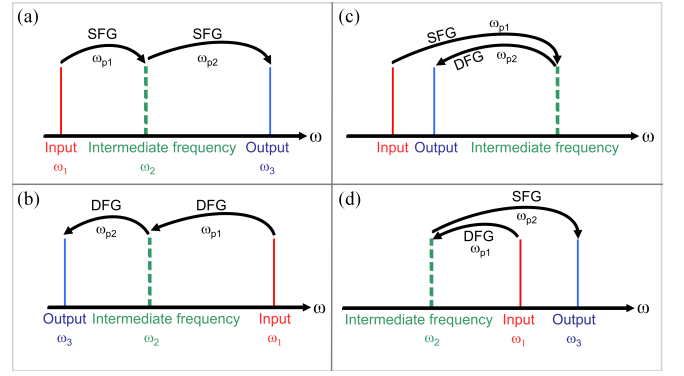}
\par\end{centering}

\caption{The four cases described by Eq. \ref{eq:CWE}: (a) two SFG processes
(b) two DFG processes (c) SFG followed by DFG (d) DFG followed by
SFG.}

\label{fig:two_step_conversion_cases}
\end{figure}

The corresponding coupled wave dynamics equations for this scenario
are

\begin{eqnarray}
i\frac{d}{dz}|\psi\rangle & = & M|\psi\rangle\label{eq:CWE}
\end{eqnarray}

where $|\psi\rangle=\left[A_{1},A_{2},A_{3}\right]^{T}$ is the state
vector, in which $\psi_{j}=A_{j}\left(z\right)$ is the complex envelope
of the amplitude of the wave with frequency $\omega_{j}$, and

\begin{equation}
M=-\left[\begin{array}{ccc}
0 & \kappa_{12}e^{-i\Delta k_{1}z} & 0\\
\kappa_{21}e^{i\Delta k_{1}z} & 0 & \kappa_{23}e^{-i\Delta k_{2}z}\\
0 & \kappa_{32}e^{i\Delta k_{2}z} & 0
\end{array}\right]\label{eq:coupling_matrix}
\end{equation}
 is the coupling matrix. Here $\kappa_{12}=\left[\chi^{\left(2\right)}\left(\omega_{1},\omega_{p_{1}};\omega_{2}\right)\omega_{1}^{2}/k_{1}c^{2}\right]\left[Re\left\{ A_{p_{1}}\right\} \mp iIm\left\{ A_{p_{1}}\right\} \right]$
and $\kappa_{23}=\left[\chi^{\left(2\right)}\left(\omega_{2},\omega_{p_{2}};\omega_{3}\right)\omega_{2}^{2}/k_{2}c^{2}\right]\left[Re\left\{ A_{p_{2}}\right\} \mp iIm\left\{ A_{p_{2}}\right\} \right]$,
where $\kappa_{ij}=\left(\omega_{i}^{2}k_{j}/\omega_{j}^{2}k_{i}\right)\kappa_{ji}^{*}$
are the effective coupling coefficients between the fields. $A_{p_{1}}$
and $A_{p_{2}}$ are the complex envelopes of the amplitudes of the
pumps fields, $\chi^{\left(2\right)}$ is the second-order nonlinear
coefficient of the material and $c$ is the velocity of light. $\Delta k_{1}=k_{1}\pm k_{p_{1}}-k_{2}$
and $\Delta k_{2}=k_{2}\pm k_{p_{2}}-k_{3}$ are the phase mismatches
of the two nonlinear processes, where $k_{j}=n_{j}\left(\omega_{j}\right)\omega_{j}/c$
is the wavenumber of the wave with frequency $\omega_{j}$. The undepleted
pumps approximation $dA_{p_{1}}/dz=dA_{p_{2}}/dz=0$ is included implicitly.
These equations can describe four different cases: two SFG processes,
two DFG processes, SFG followed by DFG and DFG followed by SFG (see
fig. \ref{fig:two_step_conversion_cases}). The difference between
these processes is manifested in the choice of either the top or bottom
sign in $\kappa_{ij}$ and $\Delta k_{j}$. The top sign corresponds
to SFG, while the bottom sign corresponds to DFG.

These equations are isomorphic to the dynamics equations of a three
level atom interacting with two EM fields \cite{Tannor_QM_book,Vitanov_ARPC_52}:
the optical interaction takes place over space (the $z$ axis in our
notation), while the atomic interaction takes place over time. The
complex amplitude of each field $A_{j}$ corresponds to the probability
amplitude of population in each atomic state, often denoted $a_{j}$.
Each effective coupling coefficient $\kappa_{ij}$ is the counterpart
of the dipole coupling strength between atomic levels $i$ and $j$,
usually described using the Rabi frequency $\Omega_{ij}\left(t\right)=d_{ij}\cdot\varepsilon\left(t\right)/\hbar$,
where $d_{ij}$ is the dipole moment between levels $i$ and $j$,
$\varepsilon\left(t\right)$ is the induced EM field, and $\hbar$
is the reduced Planck's constant. Finally, the phase mismatches, $\Delta k_{1}$
and $\Delta k_{2}$, correspond to the detunings with the two atomic
transitions, $\Delta_{1}$ and $\Delta_{2}$, respectively. We note
one minor difference: due to symmetry considerations, in the three
level atom $\Omega_{ij}=\Omega_{ji}^{*}$ when the two coupling lasers
have the same peak intensity, while in the nonlinear optics case $\kappa_{ij}=\left(\omega_{i}^{2}k_{j}/\omega_{j}^{2}k_{i}\right)\kappa_{ji}^{*}\neq\kappa_{ji}^{*}$,
as noted above.

As explained and demonstrated in previous works \cite{Suchowski_PRA_78},
these equations, which assume quasi-monochromatic laser beams for
the optical case, are applicable to pulses as short as $1ps$, owing
to the mismatch between the group velocities of the different interacting
wavelengths. Shorter pulses can be stretchered before the interaction
and compressed to their transformed limited duration after the interaction
\cite{Suchowski_APB_105}.

\subsection{Analysis of interaction under adiabatic variation}

A brief description of atomic STIRAP will be given here, while a more
rigorous analysis will be performed for the nonlinear optics variant.
In atomic STIRAP \cite{Tannor_QM_book,Vitanov_ARPC_52}, atomic population
is transferred from an initial state to a final state, where there
is no dipole coupling between these two states. Rather, each of these
two states is dipole-coupled to an intermediate state. Interestingly,
STIRAP achieves this population transfer without ever having significant
population in the intermediate state. This feat is accomplished by
adiabatically varying the Rabi frequencies $\Omega_{ij}\left(t\right)$
via control over the temporal shape of the induced EM field $\varepsilon\left(t\right)$.
Specifically, a pulse that couples the intermediate and final state
is introduced into the atomic system, followed by another pulse which
couples the intermediate state with the initial state, such that these
two pulses have some temporal overlap. Only when this counter-intuitive
pulse order is used is STIRAP achieved and the intermediate state
remains unpopulated. This absence of population in the intermediate
state renders the population transfer insensitive to radiative losses
which may result from the short life time of the intermediate state,
hence in this context it is termed a 'dark state'.

Note that single-photon resonance $\Delta_{1}=\Delta_{2}=0$ is not
required for STIRAP; two-photon resonance $\Delta_{1}+\Delta_{2}=0$
is sufficient. However, for the remainder of this section we will
consider the case of perfect phase-matching $\Delta k_{1}=\Delta k_{2}=0$,
analogous to the atomic single-photon resonance. 

An analysis of nonlinear optics STIRAP will now be given in a manner
completely analogous to the standard analysis given in the atomic
case \cite{Tannor_QM_book,Vitanov_ARPC_52}. The three eigenvalues
of the coupling matrix $M$ are

\begin{equation}
\kappa_{0}=0\,,\,\kappa_{\pm}=\pm\sqrt{\kappa_{12}\kappa_{21}+\kappa_{23}\kappa_{32}}
\end{equation}

and their corresponding normalized eigenvectors are

\begin{eqnarray}
|g_{0}\rangle & = & \frac{1}{\kappa_{c}}\left[\begin{array}{c}
\kappa_{32}\\
0\\
-\kappa_{12}
\end{array}\right]\,,\,|g_{\pm}\rangle=\frac{1}{\sqrt{\kappa_{c}^{2}+\kappa_{s}^{2}}}\left[\mp\begin{array}{c}
\kappa_{12}\\
\kappa_{s}\\
\kappa_{32}
\end{array}\right]
\end{eqnarray}

where we have defined $\kappa_{c}=\sqrt{\kappa_{12}^{2}+\kappa_{32}^{2}}$
and $\kappa_{s}=\sqrt{\kappa_{12}\kappa_{21}+\kappa_{23}\kappa_{32}}$.

Next, let us define the angle $\theta$:

\begin{equation}
\theta\equiv tan^{-1}\left(\frac{\kappa_{12}}{\kappa_{32}}\right)
\end{equation}

$|g_{0}\rangle$ can now be written as

\begin{equation}
|g_{0}\rangle=\left(\begin{array}{c}
cos\theta\\
0\\
-sin\theta
\end{array}\right)
\end{equation}

This eigenvector has an important property: it has no contribution
from the intermediate state, i.e. $\langle g_{0}|A_{2}\rangle=0$.
It is this property which will enable energy to be transferred from
$A_{1}$ to $A_{3}$ without going through $A_{2}$, as will now be
explained.

We assume that initially all of the optical power is in $A_{1}$,
and additionally $\left|\kappa_{12}\right|,\left|\kappa_{21}\right|\ll\left|\kappa_{23}\right|,\left|\kappa_{32}\right|$,
so $\theta\approx0$. In this situation, out of the three eigenstates,
only $|g_{0}\rangle$ has a nonzero term, thus the system can be said
to be in this eigenstate. During the interaction, the coupling coefficients
change adiabatically, such that at its end $\left|\kappa_{12}\right|,\left|\kappa_{21}\right|\gg\left|\kappa_{23}\right|,\left|\kappa_{32}\right|$,
so $\theta\approx\pi/2$. According to the adiabatic theorem, if the
system is at an eigenstate and is subject only to adiabatic changes,
it will remain at the same eigenstate, in this case $|g_{0}\rangle$.
However, when $\theta\approx\pi/2$, the physical meaning of $|g_{0}\rangle$
is that all of the optical power is in $A_{3}$. We therefore conclude
that the system will remain in the $|g_{0}\rangle$ eigenstate throughout
the interaction, going from the initial state $\left[1,0,0\right]^{T}$
(all power in $A_{1}$) to the final state $\left[0,0,-1\right]^{T}$
(all power in $A_{3}$). Since $|g_{0}\rangle$ contains no component
of the intermediate frequency amplitude, this means that power will
be transferred from $A_{1}$ to $A_{3}$ without ever going through
$A_{2}$. Note that the counter-intuitive order is maintained in the
nonlinear optics case: first $A_{2}$ and $A_{3}$ are coupled, and
the coupling between $A_{1}$ and $A_{2}$ is introduced at a later
point.

In order to satisfy the adiabatic condition, it is required that the
coupling between the desired eigenstate, $|g_{0}\rangle$, and the
other eigenstates, $|g_{\pm}\rangle$, is small compared to the difference
between the effective wavenumbers of these states \cite{Vitanov_ARPC_52}:

\begin{equation}
\left|\langle\frac{dg_{0}}{dz}|g_{\pm}\rangle\right|\ll\left|\kappa_{0}-\kappa_{\pm}\right|
\end{equation}

This condition can be written in terms of $\theta$ as

\begin{equation}
\left|\frac{d\theta}{dz}\right|\ll\frac{\kappa_{s}}{\kappa_{c}\sqrt{\kappa_{c}^{2}+\kappa_{s}^{2}}}\label{eq:adiab_cond}
\end{equation}

where $\theta$ is the adiabatically varying parameter, going from
$0$ to $\pi/2$.

\begin{figure}
\begin{centering}
\includegraphics[width=1\columnwidth]{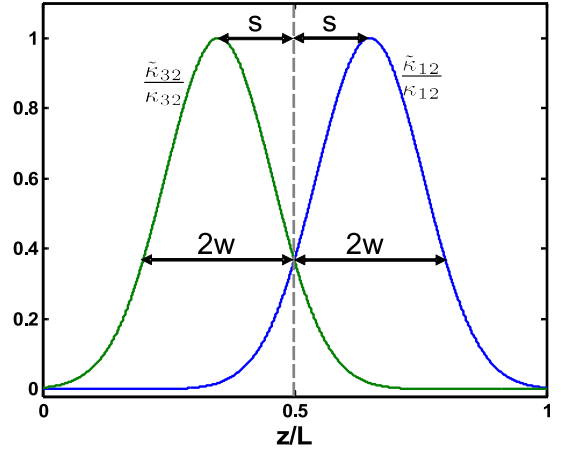}
\par\end{centering}

\caption{Normalized coupling coefficients of the two nonlinear processes with
Gaussian modulation. Note that the $\omega_{2}\leftrightarrow\omega_{3}$
coupling is (counter-intuitively) maximized before that of $\omega_{1}\leftrightarrow\omega_{2}$.}

\label{fig:mod_gauss}
\end{figure}

Optimization of the atomic STIRAP process by temporally shaping $\Omega_{ij}\left(t\right)$,
and also using time varying detunings $\Delta_{j}\left(t\right)$,
has been studied extensively \cite{Laine_PRA_53,Vasilev_PRA_80,Dridi_PRA_80,Xi_PRL_105}.
Here we will use Gaussian modulation of the nonlinear coupling in
space (along the interaction), which is convenient for analysis and
illustrates the main features of STIRAP. For this purpose we replace
the coupling coefficients in the matrix $M$ of Eq. \ref{eq:coupling_matrix}
with

\begin{eqnarray}
\tilde{\kappa}_{12}\left(z\right) & = & \kappa_{12}e^{-\left(z-L/2-s\right)^{2}/w^{2}}\nonumber \\
\tilde{\kappa}_{32}\left(z\right) & = & \kappa_{32}e^{-\left(z-L/2+s\right)^{2}/w^{2}}\label{eq:kappa_gauss_mod}
\end{eqnarray}

where $s$ and $w$ are parameters which determine the locations of
coupling maxima and the rate of the coupling variation, respectively.
See fig. \ref{fig:mod_gauss} for an illustration of these modulation
functions. Note that the $s$ parameter determines the coupling order:
for $s<0$ the intuitive order is obtained, while for $s>0$ the modulation
is in the counter-intuitive order regime. Additionally, as before,
$\tilde{\kappa}_{ij}=\left(\omega_{i}^{2}k_{j}/\omega_{j}^{2}k_{i}\right)\tilde{\kappa}_{ji}^{*}$.

\section{Numerical Simulation Under Ideal conditions\label{sec:Numerical-Simulation-Ideal}}

\subsection{Interaction in the Absence of Absorption}

In this section the main features of optical STIRAP will be demonstrated
by numerical simulation, which solves Eq. \ref{eq:CWE}. For this
purpose, it will be assumed that both processes are simultaneously
phase-matched, and that the desired modulation of the coupling coefficient
$\tilde{\kappa}_{ij}$ has been achieved. In this respect, the conditions
are considered ideal. Section \ref{sec:Numerical-Simulation-PRQPM}
will discuss a technique to imitate this situation using currently
available technology.

\begin{figure}
\begin{centering}
\includegraphics[width=1\columnwidth]{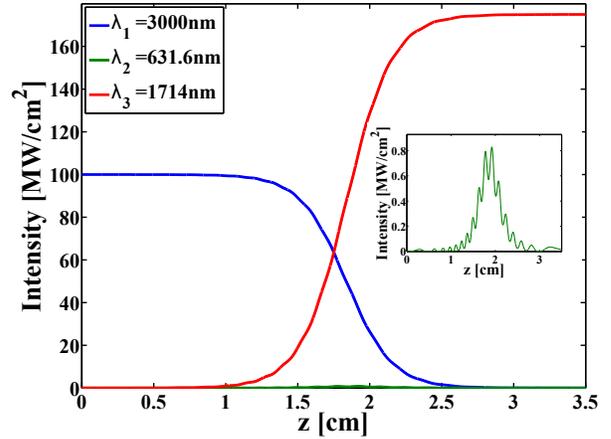}
\par\end{centering}

\caption{Numerical simulation of the intensities of the interacting waves along
the nonlinear medium, with no intermediate wave absorption, under
ideal conditions and counter-intuitive order modulation ($s=5mm$).
The inset shows the intermediate wave intensity on a smaller scale.}

\label{fig: I_vs_z}
\end{figure}

In the simulation, the two processes were SFG and DFG: $\left(\lambda_{1}=3000nm\right)+\left(\lambda_{p_{1}}=800nm\right)\rightarrow\left(\lambda_{2}=631.6nm\right)$
and $\left(\lambda_{2}=631nm\right)-\left(\lambda_{p_{2}}=1000nm\right)\rightarrow\left(\lambda_{3}=1714nm\right)$.
The nonlinear medium was a 35mm long $KTiOPO_{4}$ (KTP) crystal with
$\chi^{\left(2\right)}\left(\omega_{1},\omega_{p_{1}};\omega_{2}\right)\approx\chi^{\left(2\right)}\left(\omega_{2},\omega_{p_{2}};\omega_{3}\right)=16.65pm/V$
\cite{Reshak_JPCB_114}. The coupling coefficients refractive indexes
were calculated using the Sellmeier equation of Fradkin et. al \cite{Fradkin_APL_74}
and Emanueli et. al \cite{Emanueli_AO_42} at a temperature of $100^{o}C$.
The modulation parameters were chosen to be $s=5mm$ and $w=8mm$.
Note that for every $s>0$ the interaction was performed using the
counter-intuitive order. The input intensity was $100MW/cm^{2}$ and
each of the two pumps had an intensity of $2GW/cm^{2}$. Figure \ref{fig: I_vs_z}
shows the resulting intensities of the interacting waves along the
nonlinear crystal, with the intermediate wave intensity on a smaller
scale in the inset. The input power is seen to be fully converted
to the output wave, with some additional power from the first pump:
if every photon at $\lambda_{1}$ is converted to a photon at $\lambda_{3}$,
the intensity ratio is $I_{3}/I_{1}=\omega_{3}/\omega_{1}=\lambda_{1}/\lambda_{3}$.
For the parameters used here, full conversion means $I_{3}=175MW/cm^{2}$.
Furthermore, the intermediate power is at most $0.8\%$ of the input
power. These are exactly the characteristics of STIRAP.

Interestingly, using the method of adiabatic elimination \cite{Porat_OE_20}
for the same set of parameters and under the same ideal conditions,
the output intensity would reach only $73.5MW/cm^{2}$. Clearly, the
STIRAP analog method introduced here is more efficient, while maintaining
the property of negligible intermediate frequency power.

\begin{figure}
\begin{centering}
\includegraphics[width=1\columnwidth]{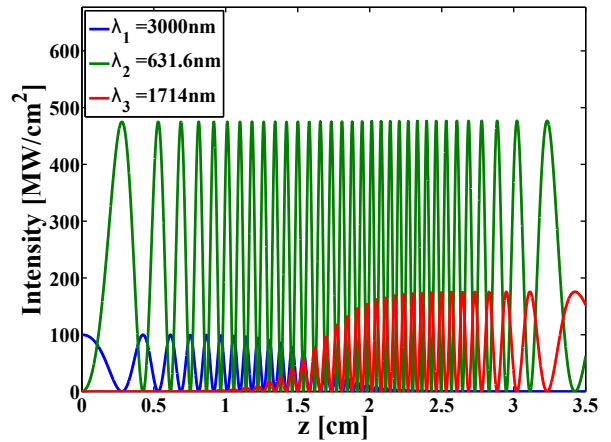}
\par\end{centering}

\caption{Numerical simulation of the intensities of the interacting waves along
the nonlinear medium, with no intermediate wave absorption, under
ideal conditions and intuitive order modulation ($s=-5mm$).}

\label{fig: I_vs_shift_intuitive}
\end{figure}

For comparison, we repeated the same simulation with $s=-5mm$, i.e.
with intuitive order. The results are depicted in fig. \ref{fig: I_vs_shift_intuitive},
showing strong oscillations of all three waves intensities. Specifically,
significant energy is present in the intermediate frequency, in contrast
to the counter-intuitive result of fig. \ref{fig: I_vs_z}.

Furthermore, we explored the conversion process dynamics as a function
of the shift parameter $s$. In order to do this, the width parameter
was kept constant at $w=8mm$ while the shift parameter was varied
from $-20mm$ to $+20mm$. Efficient conversion from $\lambda_{1}$
to $\lambda_{3}$ together with negligible ($<1\%$) power at $\lambda_{2}$
was obtained for $3.3mm<s<6.6mm$. In this shift interval, which is
in the counter-intuitive regime, the overlap between the modulation
Gaussians of Eq. \ref{eq:kappa_gauss_mod} is significant enough to
facilitate STIRAP. It was also found that for values of $s<0$, i.e.
in the intuitive regime, there is always high intermediate power somewhere
along the crystal.

\subsection{Interaction in the Presence of Intermediate Frequency Absorption}

\begin{figure}
\begin{centering}
\includegraphics[width=1\columnwidth]{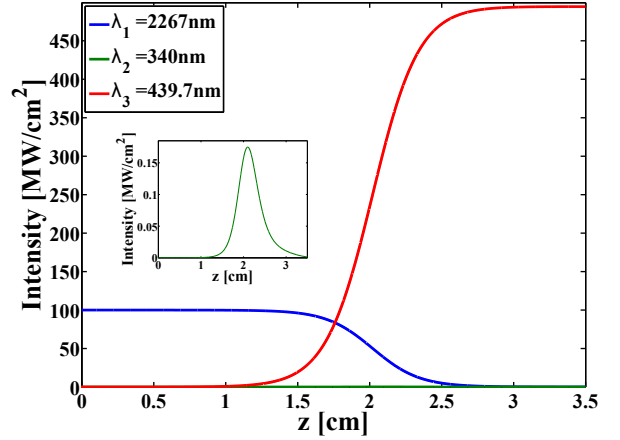}
\par\end{centering}

\caption{Numerical simulation of the intensities of the interacting waves along
the nonlinear medium, with high intermediate wave absorption and under
ideal conditions. The inset shows the intermediate wave intensity
on a smaller scale.}

\label{fig:I_vs_z_UV}
\end{figure}

One of the key advantages of having a dark intermediate state is that
the crystal absorption in the intermediate wavelength does not change
the conversion efficiency \cite{Porat_OE_20}. Here we consider the
same interaction as the previous one, except that the intermediate
frequency is in the ultraviolet absorption band of the KTP crystal.
The two processes are $\left(\lambda_{1}=2267nm\right)+\left(\lambda_{p_{1}}=400nm\right)\rightarrow\left(\lambda_{2}=340nm\right)$
and $\left(\lambda_{2}=340nm\right)-\left(\lambda_{p_{2}}=1500nm\right)\rightarrow\left(\lambda_{3}=439.7nm\right)$.
The absorption coefficient at the intermediate wavelength is $\alpha_{2}=229.9cm^{-1}$
\cite{Dudelzak_JAP_87}. Absorption is known to be accompanied by
enhancement of the nonlinear susceptibility, and this is the case
here as well. However, this enhancement is not very high, since only
one wavelength ($\lambda_{2}$) in each interaction is near an electronic
transition resonance of the KTP crystal. We can calculate this enhancement
using the density matrix quantum mechanical model of the linear and
nonlinear susceptibilities \cite{Boyd_NLO_book}, which yields

\begin{eqnarray}
\chi^{\left(1\right)}\left(\omega\right) & \propto & \left(\omega^{\prime}-\omega-i\gamma^{\prime}\right)^{-1}\label{eq:chi1_res}\\
\chi^{\left(2\right)}\left(\omega_{j},\omega_{k};\omega_{j}+\omega_{k}\right) & \propto & \left(\omega^{\prime}-\omega_{j}-\omega_{k}-i\gamma^{\prime}\right)^{-1}\times\nonumber \\
 &  & \left[\left(\omega^{\prime\prime}-\omega_{j}-i\gamma^{\prime\prime}\right)^{-1}\right.+\nonumber \\
 &  & \left.\left(\omega^{\prime\prime\prime}-\omega_{k}-i\gamma^{\prime\prime\prime}\right)^{-1}\right]\label{eq:chi2_res}
\end{eqnarray}
where the primed $\omega$ and $\gamma$ terms are resonant transition
frequencies and full widths at half maximum, respectively, and $\omega_{j}$
and $\omega_{k}$ are the two low frequencies out of the three in
the TWM process. In the case under study, each of the two low frequencies
in each of the two processes, namely $\omega_{1},\omega_{p1},\omega_{p2}$
and $\omega_{3}$, are far from any resonance. The last two terms
of Eq. \ref{eq:chi2_res} therefore do not contribute to the resonant
enhancement of $\chi^{\left(2\right)}$. The enhancement stems only
from the first term, since it refers to the high frequency which is
the intermediate frequency $\omega_{2}$ in each of the two processes
considered here. From Eq. \ref{eq:chi1_res} and \ref{eq:chi2_res}
is it therefore clear that in this case the second order susceptibility
scales like the first order susceptibility. The enhancement factor
for the wavelengths involved here can be calculated using the relation
between the linear susceptibility and the refractive index, by making
use of the Sellmeier equation of KTP:

\begin{alignat}{1}
\frac{\chi^{\left(2\right)}\left(\lambda_{1}+\lambda_{p1}\rightarrow\lambda_{2}\right)}{\chi^{\left(2\right)}\left(1064nm+1064nm\rightarrow532nm\right)} & =\nonumber \\
\frac{\chi^{\left(2\right)}\left(\lambda_{2}-\lambda_{p2}\rightarrow\lambda_{3}\right)}{\chi^{\left(2\right)}\left(1064nm+1064nm\rightarrow532nm\right)} & =\nonumber \\
\frac{\chi^{\left(1\right)}\left(\lambda_{2}\right)}{\chi^{\left(1\right)}\left(1064nm\right)}=\frac{n^{2}\left(\lambda_{2}\right)-1}{n^{2}\left(1064nm\right)-1} & =1.4
\end{alignat}
where we have used the $1064nm$ second harmonic generation value
of $\chi^{\left(2\right)}$ as a reference. This enhancement is equivalent
to increasing the intensity of each of the two pumps by a factor of
$1.4^{2}\approx2$. Using these parameters together with the same
input intensity, pump intensities and Gaussian modulation of the coupling
coefficients as before, Eq. \ref{eq:CWE} was solved once again. The
resulting intensities along the crystal are displayed in fig. \ref{fig:I_vs_z_UV},
which shows that the conversion from $\lambda_{1}$ to $\lambda_{3}$
is still efficient (here full conversion means $I_{3}=515MW/cm^{2}$
due to contribution from the first pump) . The intermediate wave intensity,
displayed in the inset, reaches at most $0.17\%$ of the input intensity.
These results reveal the true merit of avoiding significant power
at the intermediate frequency, as it allows the conversion to remain
efficient in spite of great absorption at $\lambda_{2}$.

\section{Numerical Simulation Using Phase-Reversal Quasi-Phase-Matching\label{sec:Numerical-Simulation-PRQPM}}

In the preceding section it was assumed that both processes were phase
matched and that the coupling coefficients were modulated as desired,
without accounting for any mechanism to achieve these purposes. Now
these assumptions will not be made. Instead, a technique called phase-reversal
quasi-phase matching (PRQPM) \cite{Chou_OL_24} will be employed to
provide phase-matching as well as to modulate the coupling coefficients.
In PRQPM, binary functions are multiplied to produce a desired modulation,
which has multiple Fourier components. This modulation can therefore
provide QPM for multiple processes, i.e. processes with different
phase-mismatches. For example, we can construct a product of two binary
functions as follows:

\begin{eqnarray}
g\left(z\right) & = & sign\left[-cos\left(\pi D_{1}\right)+cos\left(\frac{2\pi}{\Lambda_{1}}z\right)\right]\times\nonumber \\
 &  & sign\left[-cos\left(\pi D_{2}\right)+cos\left(\frac{2\pi}{\Lambda_{2}}z\right)\right]\label{eq:PRQPM_mod}
\end{eqnarray}

For each periodic term in the product, $D_{j}$ is the duty cycle
and $\Lambda_{j}$ is the period. As always in QPM, we keep only the
first order Fourier terms that contribute the desired phase-mismatches.
If we choose $\Lambda_{1}=\Delta k_{1}$ and $\Lambda_{2}=\Delta k_{2}$,
using simple Fourier analysis yields

\begin{eqnarray}
g_{QPM}\left(z\right) & \approx & \frac{2}{\pi}\left(2D_{2}-1\right)sin\left(\pi D_{1}\right)exp\left(\pm i\Delta k_{1}\right)+\nonumber \\
 &  & \frac{2}{\pi}\left(2D_{1}-1\right)sin\left(\pi D_{2}\right)exp\left(\pm i\Delta k_{2}\right)\label{eq:PR_QPM_mod_g}
\end{eqnarray}

Evidently, this choice of the modulation periods provides QPM for
the desired processes, while $D_{1}$ and $D_{2}$ determine the the
magnitude of the effective coupling coefficient for each process.
Varying the two duty cycles along the crystal achieves the required
modulation.

The two cases regarded in section \ref{sec:Numerical-Simulation-Ideal},
with ideal phase-matching and coefficient modulation, are considered
again in this section. In addition to using PRQPM for phase-matching
and coefficient modulation, the calculation made here also takes into
consideration technological restrictions on nonlinear modulation,
and pumps depletion/amplification. We shall now assume that QPM is
achieved by the wide-spread technique of electric field poling of
a ferroelectric crystal.

\begin{figure}
\begin{centering}
\includegraphics[width=1\columnwidth]{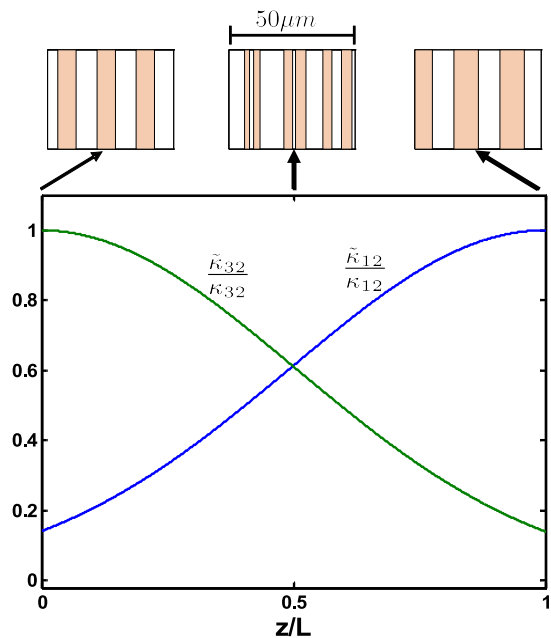}
\par\end{centering}

\caption{Normalized coupling coefficients of the two nonlinear processes with
the modulation used in section \ref{sec:Numerical-Simulation-PRQPM}.
The top panels show the PRQPM poling, which was used in the simulation
of fig \ref{fig:I_vs_z_PRQPM}, at the beginning, center and end of
the nonlinear crystal. The colored and white stripes represent domains
with positive and negative nonlinearity.}

\label{fig:mod_gauss_PRQPM}
\end{figure}

Using QPM, especially when considering the technological limitations
of this technique, makes it more difficult to satisfy the adiabaticity
condition of Eq. \ref{eq:adiab_cond}, which essentially requires
the changes in the coefficients to be very gradual. For this purpose,
we first note that the spatial area where $\tilde{\kappa}_{32}$ is
rising while $\tilde{\kappa}_{12}\approx0$ is going to waste, since
$\theta=0$ all along it (see fig. \ref{fig:mod_gauss}). In order
to achieve adiabatic interaction, it is sufficient that at the start
$\tilde{\kappa}_{12}\ll\tilde{\kappa}_{32}$. We can therefore start
the interaction with $\tilde{\kappa}_{32}\left(0\right)=max\left[\tilde{\kappa}_{32}\left(z\right)\right]$,
i.e. by choosing $s=L/2$, and use the additional space, which was
previously wasted, to make the coefficients gradients smaller, by
increasing the modulation width $w$. For this reason, in the simulations
conducted using PRQPM, the shift parameter was $s=L/2=17.5mm$ and
the width parameter was $w=25mm$. The resulting normalized coefficients
are displayed in fig. \ref{fig:mod_gauss_PRQPM}, along with an example
of a corresponding PRQPM poling pattern at various locations.

Furthermore, we note that Zukauskas et. al succeeded in fabricating
domains as small as $345nm$ in KTP \cite{Zukauskas_OME_1}. In the
simulation conducted here, domains formed by Eq. \ref{eq:PRQPM_mod},
which were less than $350nm$ long, were concatenated with neighboring
domains until they reached this minimum length. In this manner modulation
patterns that are attainable with current technology have been constructed.

\begin{figure}
\begin{centering}
\includegraphics[width=1\columnwidth]{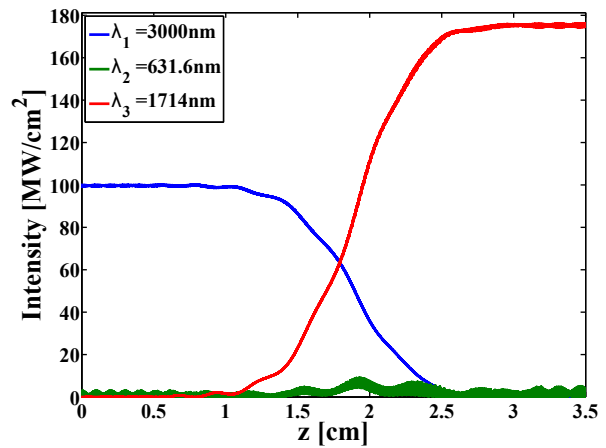}
\par\end{centering}

\caption{Numerical simulation of the intensities of the interacting waves along
the nonlinear medium, with no intermediate wave absorption and where
the crystal is modulated using phase-reversal quasi-phase-matching.}

\label{fig:I_vs_z_PRQPM}
\end{figure}

Simulation results for the first set of wavelengths, all of which
are in the crystal's transparency spectral region, are displayed in
fig. \ref{fig:I_vs_z_PRQPM}. Comparing with the ideal case results
shown in fig. \ref{fig: I_vs_z}, we see that the main features of
the interaction have been preserved: high conversion from $\lambda_{1}$
to $\lambda_{3}$ is obtained, while the intermediate wavelength intensity
remains relatively low. Still, there are two clear differences from
the ideal case: the intermediate intensity reaches a higher value
($9.15\%$ of the input) and the intensities of all three waves fluctuate
along the crystal. Both of these differences come from the use of
PRQPM and the domain size limit: the oscillations are a well established
property of QPM, coming from the non-phase-matched part of the interaction,
i.e. higher Fourier orders of the modulation. The desired adiabatic
variation is obtained on an average scale, which includes many domains.
As a result, on a shorter scale, energy does get transferred to the
intermediate wave. The modulation periods being $\Lambda_{1}=18.2\mu m$
and $\Lambda_{1}=15.5\mu m$, the domain size limit makes it more
difficult to satisfy the adiabatic condition, resulting in more energy
transfer to $\lambda_{2}$.

\begin{figure}
\begin{centering}
\includegraphics[width=1\columnwidth]{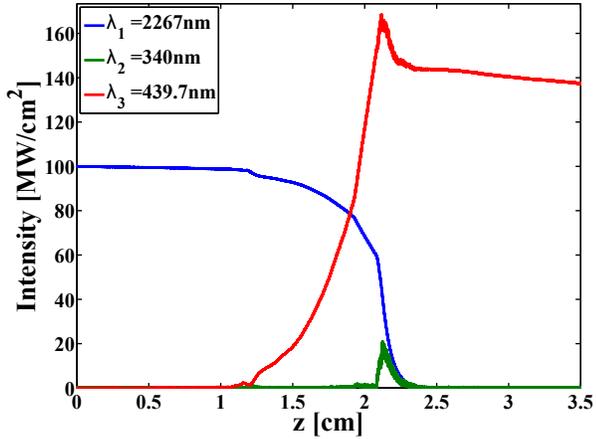}
\par\end{centering}

\caption{Numerical simulation of the intensities of the interacting waves along
the nonlinear medium, with high intermediate wave absorption and where
the crystal is modulated using phase-reversal quasi-phase-matching.}

\label{I_vs_z_UV_PRQPM}
\end{figure}

Simulation results for the second set of wavelengths, which include
a highly absorbed intermediate wave, are depicted in fig. \ref{I_vs_z_UV_PRQPM}.
While significant conversion from $\lambda_{1}$ to $\lambda_{3}$
took place in this case, it has some features which clearly distinguish
it from the ideal case, in addition to the usual QPM oscillations
noted above. First, the intermediate intensity reaches a value as
high as $21.1\%$ of the input intensity, where significant intermediate
power is only present in a confined spatial area. The reason for this
phenomenon is that in this area the domain size limitation causes
loss of adiabaticity, resulting in energy transfer from $\lambda_{3}$
to $\lambda_{2}$, which is quickly absorbed by the crystal. Note
that the modulation periods are $\Lambda_{1}=2.8\mu m$ and $\Lambda_{2}=2.1\mu m$,
so small domains are abundant. Due to the loss of adiabaticity, the
remainder of the interaction is also characterized by conversion from
$\lambda_{3}$ to $\lambda_{2}$, albeit at a slower rate since little
energy is present at the intermediate frequency at any point, due
to the strong absorption.

Optical STIRAP via PRQPM, in the presence of intermediate frequency
absorption, is ultimately limited by the ratio between the absorption
length $l_{abs}=1/\alpha_{2}$ and the coherence lengths $l_{c}^{\left(j\right)}=\pi/\Delta k_{j}\,,\, j=1,2$
\cite{Boyd_NLO_book}. In order for any QPM scheme to have a significant
effect, it is required that $l_{abs}>l_{c}$, otherwise an interacting
wave will be absorbed before being affected by the modulation (for
the case considered here, we have $l_{abs}/l_{c}^{\left(1\right)}=31.9$
and $l_{abs}/l_{c}^{\left(2\right)}=41.1$). Nevertheless, optical
STIRAP via PRQPM still greatly reduces the conversion efficiency's
sensitivity to intermediate frequency absorption, since considerably
less energy is ever present in the intermediate wave. Furthermore,
no critical control of the modulation is required, as is the case
for non-adiabatic QPM methods in the presence of absorption, which
significantly limits the interaction length.

An additional limitation of this modulation scheme is that in order
to achieve independent control of each of the two processes via the
duty cycles $D_{1}$ and $D_{2}$, their phase-mismatches need to
be sufficiently different from one another. Otherwise, each of the
two terms in Eq. \ref{eq:PR_QPM_mod_g} will have the same effect
on both processes, and independent control would be lost.

\section{Future Prospects and Applications\label{sec:Future-Prospects-and}}

In the previous section limitations of PRQPM in face of current fabrication
capabilities were discussed. Specifically, it was demonstrated that
the minimum domain size limits this method's capability of satisfying
the adiabaticity condition when the dispersion is stronger, i.e. when
the interacting wavelengths are relatively short (in the near UV).
We consider it reasonable to assume that the current trend of ever
decreasing available domain size will continue \cite{Zukauskas_OME_1},
making PRQPM feasible to processes involving short wavelengths, going
into the UV. Note also that the domain size limitation can be somewhat
circumvented, by relinquishing single photon resonance $\Delta k_{1}=\Delta k_{2}=0$
and satisfying only two photon resonance $\Delta k_{1}+\Delta k_{2}=0$.
However, this will come at a cost of efficiency.

Moreover, conversion efficiency can be further improved by employing
techniques originally invented for atomic STIRAP \cite{Laine_PRA_53,Vasilev_PRA_80,Dridi_PRA_80,Xi_PRL_105}.
In these techniques the temporal dependence of the Rabi frequencies
and detunings are controlled by pulse shaping and chirping mechanisms,
respectively. Both of these controls have counter-parts in PRQPM optical
STIRAP. As was shown here, the coupling coefficients can be controlled
via the duty cycles of the PRQPM modulation function of Eq. \ref{eq:PR_QPM_mod_g},
in analogy with pulse shaping of the Rabi frequencies. Additionally,
the effective phase-mismatch of each process can be modulated along
the interaction by variations in the periods of the modulation ($\Lambda_{1}$
and $\Lambda_{2}$ in Eq. \ref{eq:PR_QPM_mod_g}). This is the counterpart
of variable atomic detunings, obtained by chirping.

An entirely different method for obtaining adiabatic interaction can
involve anomalous dispersion phase-matching (ADPM) and tight beam
focusing. In ADPM, absorption resonances are deliberately brought
forth, either by doping bulk nonlinear crystals with metals or ions
\cite{Anderson_CM_8}, or by introducing chromophores in the fabrication
process of polymer waveguides \cite{Dietrich_CPL_280}. These resonances
influence the dispersion properties of the nonlinear medium, thus
enabling birefringent phase-matching of various processes. Nonlinear
optics STIRAP can be facilitated by using ADPM for phase-matching,
while focusing the two pump beams at different locations along the
crystal, achieving the modulation of the coupling coefficients simply
by allowing diffraction to modulate pumps intensities (working with
focused beams would invalidate the plane-wave approximation used in
this work, however we believe that this would not prevent our method
from being applied, and that it can be compensated by increasing pump
power). Furthermore, the intermediate frequency can reside within
the absorption band of the doped crystal without hindering the conversion
efficiency.

Two applications of nonlinear optics STIRAP will now be put forth.
First, a straightforward application would simply be frequency conversion
across large frequency ranges. By employing two SFG or two DFG processes,
efficient conversion can be performed where the input and output frequencies
are greatly separated from each other (e.g. mid-IR input and blue
visible output, or vice versa). Second, we suggest that this method
can be employed in the study of nonlinear crystal electronic structure
and properties, by employing a SFG process followed by a DFG process.
While Raman spectroscopy enables the study of molecular structure
in the infrared absorption band, experimental data on the electronic
structure in the UV absorption band is scarce \cite{Reshak_JPCB_114,Cabuk_CEJP_10}.
Experiments to this affect rely mostly on nonlinear absorption and
refraction \cite{DeSalvo_IEEE_JQE_32,Li_OC_144}, and are limited
by linear absorption. UV electronic resonances will have a significant
influence on nonlinear optics STIRAP processes, first via phase-matching
and second via enhancement of the nonlinear coefficient. For example,
by using a tunable pump or by tuning the crystal temperature, phase
matching conditions can be controlled, thus allowing direct measurement
of dispersion inside the UV absorption band (i.e. beyond the band
gap energy).

\section{Conclusion}

This work theoretically demonstrates a method for efficient frequency
conversion through an intermediate frequency, which never receives
any significant amount of power throughout the interaction. This property
was predicted analytically by analogy with atomic STIRAP, and confirmed
with numerical simulations. It was shown that the absence of power
at the intermediate frequency renders the conversion efficiency highly
insensitive to absorption of the intermediate frequency, opening the
way to conversion through absorptive bands in the UV. A technologically
feasible method for carrying out such a process was proposed and numerically
demonstrated to be effective. We suggest to apply this method in conversion
between highly disparate frequencies, e.g. mid-IR to visible conversion.
Another application is the study of the electronic structure and properties
of nonlinear optical crystals, which is manifested in the UV absorption
band, now made accessible in spite of linear absorption.
\begin{acknowledgments}
The authors would like to acknowledge Dr. Ori Katz for suggesting
the resonant enhancement of the nonlinear susceptibility. This work
was supported by the Israel Science Foundation\end{acknowledgments}

\end{document}